\documentclass[%
 reprint,
superscriptaddress,
showpacs,preprintnumbers,
 amsmath,amssymb,
 aps,
]{revtex4-1}

\usepackage{graphicx}% Include figure files
\usepackage{dcolumn}% Align table columns on decimal point
\usepackage{bm}% bold math

\usepackage{color}
\usepackage[colorlinks=true,citecolor=blue]{hyperref}
\hypersetup{colorlinks=true,citecolor=blue,linkcolor=red
,urlcolor=blue}

\begin{document}

\title{Negative differential thermal conductance in a borophane normal metal-superconductor junction }

\author{Moslem Zare}
\email{mzare@yu.ac.ir}
\affiliation{Department of Physics, Yasouj University, 75914-353, Yasouj, Iran}
\begin{abstract}
We study the charge and heat transport in a normal-metal/superconductor (NS) junction of the tilted anisotropic Dirac cone material borophane, using the extended Blonder-Tinkham-Klapwijk formalism. We numerically find that, in spite of the large mismatch in the Fermi wavevectores of the normal-metal and superconductor sides of the borophane NS junction, the electron-hole conversion happens with unit probability at normal incidences.
Furthermore, in the heavily doped superconducting regime, for heavily doped normal borophane, the electron-hole conversion happens with unit probability, almost at any incident angle.
The dependence of the Andreev conductance on the Fermi energy and bias voltage, enable us selecting the retro or specular type of Andreev reflection process.
We numerically find that, independent of the Fermi energy, the temperature dependence of the thermal conductance in borophane can be modelled as an inverse Gaussian function, reflecting the d-wave symmetry of the borophane superconductor.
We propose a scheme for achieving negative differential thermal conductance at intermediate Fermi energies.
Our findings will have potential applications in developing borophane-based thermal management and signal manipulation mesoscopic structures such as heat transistors, heat diodes and thermal logic gates.
\end{abstract}


\maketitle
\section{Introduction}\label{sec:intro}
The Andreev reflection (AR) is a scattering process at the interface between a normal-metal conductor (N) and a superconductor (S), in which an incident electron from the N side is reflected as a hole, and the created Cooper pair moves through the superconductor transferring a charge 2e. The electron-hole conversion occurs at the excitation energies $\varepsilon$, less than the superconducting energy gap $\Delta_S$ ($\varepsilon<\Delta_S$) ~\cite{Andreev64}.
In this process, known as retro Andreev reflection (RAR), an electron with an energy $\varepsilon$ (relative to the Fermi energy $E_F$), a momentum $k_F+\delta k$ and spin polarization $s$, upon hitting the N/S interface, is retro reflected as a hole of opposite spin $-s$, with energy $-\varepsilon$ and momentum $-k_F+\delta k$, results in a finite conductance in the N/S junction ~\cite{BTK}.
The occurrence of the AR has been shown, e.g., by tunneling experiments~\cite{Tomasch65,Rowell73}, Gantmakher-resonance experiments~\cite{Krylov80} or by a double-point contact electron-focusing technique~\cite{Benistant-prl83}. In the retro-AR both electron and hole are located in the same conduction or valence band (the electron-hole conversion is intraband)~\cite{BTK}.

Another novel AR phenomenon, known as specular Andreev reflection (SAR), has been discovered by Beenakker~\cite{AR-G1,AR-G2} in graphene-based superconducting hybrid structures, in which the hole is reflected along a specular path of the incident electron, resulted from the sublattice pseudospin degree of freedom of electrons with a gap (non-superconducting) in its band dispersion~\cite{majidi12,majidi13}, which is absent in ordinary metal-superconductor interfaces. In the specular (retro) AR the incident electron and reflected hole are located at different (same) bands, namely interband (intraband) conversion.

In recent years, hybrid normal metal–superconductor structures have received considerable attention due to their novel properties arising from exotic transport behaviors, when plenty of new materials with massless Dirac/Weyl excitations or nontrivial topological features are discovered and used to fabricate these interfaces \cite{AR-G1,AR-G2,AR-G3,AR-G4,AR-N,AR-IT1,AR-TS1,AR-WSM1}.

Recent advances in fabrication technologies have made exploring two-dimensional materials possible for applications, which in turn has triggered a tremendous interest. Transport properties of graphene-based normal-metal/superconductor heterojunctions have been investigated extensively and many striking properties have been obtained~\cite{AR-G1, sengupta_prl_06,Ludwig_prb_07, linder_prl_07}.
\par
In the case of specular Andreev reflection for undoped graphene $(E_\text{F}=0)$, holes are produced in the valence band. In doped graphene ($E_\text{F} > 0$), the Andreev reflection can be normal or specular, depending on the energy of the incoming electron. In heavily doped graphene
($E_\text{F} \gg \Delta_S)$, only normal Andreev reflection is present for subgap energies, since the distance from Fermi level to the valence band is too large for specular AR to occur. In an undoped graphene, Andreev reflection is an interband scattering at all excitation energies, which is not
possible in conventional metals, with no excitation gap separating the conduction and valence bands.
In the regime $0<E_\text{F}<\Delta_S$, depending on the incident electron energy $\varepsilon$, the AR has a form of either normal or specular.

The tunneling conductance~\cite{sengupta_prl_06,linder_prl_07} and Josephson current~\cite{Maiti} in graphene junctions are oscillatory functions of the width and height of the barrier at the interface.
Perfect AR has been also proposed in the NS junction of topological insulator~\cite{perfect AR1,perfect AR2}.

It is known that, in conventional NS junctions the electric and thermal conductances reflect the magnitude or symmetry of the superconductor energy gap ~\cite{Andreev64,BTK}.
Compared to the linear temperature dependence in a bulk superconductor, explained from the Wiedemann-Franz law, its thermal conductivity is exponentially suppressed at the low temperatures $k_B T\ll\Delta _S$  \cite{Bardeen}, because Andreev reflection completely blocks the subgap flow of energy and thus their thermal conductivity is often negligible. The story is considerably different in hybrid mesoscopic structures of NS because the heat flow through superconductor may becomes significant \cite{Courtois}.
Applying an electric current in the normal-metal/superconducting junction, is used foe refrigeration of electrons in the normal metal which can be used for the realization of microcoolers~\cite{Giazotto2006, Review_2, Review_3}, high-sensitive detectors and quantum devices~\cite{Clark, Ullom}.

While charge and thermal conductance in normal/superconductor junction of the isotropic Dirac materials have been studied~\cite{AR-G1,sengupta_prl_06,Ludwig_prb_07,linder_prl_07,linder_prb_14,li_prb_16,li_prb_16b,zhou_prb_16,li_prb_16c,kuzmanovski_prb_16,paul_prb_16,Maiti,
perfect AR1,Ren13,moslem-physicac}, but to our knowledge, thermal Andreev reflection in a zero-gap semimetal, with tilted anisotropic Dirac cones, has not yet been reported in the literature.
Tilted Dirac cones have been predicted in a series of materials, including deformed graphene \cite{Goerbig08,Choi10}, partially hydrogenated graphene \cite{Lu16}, the surface of topological crystalline insulators such as the (001) surface of SnSe \cite{Tanaka12, Sodemann15}, organic compound $\alpha$-(BEDT-TTF)$_{2}$I$_{3}$ \cite{Goerbig08, Kobayashi07, Hirata16, Hirata17}, 8-$Pmmn$
borophene \cite{Zhou14, Zobolotskiy16, Feng17, Sadhukhan17, Verma17,Islam17}.

Graphene-like two-dimensional (2D) structure of boron, as carbon’s neighbor in the periodic table, known as borophene has attracted great attention due to its fascinating properties and promising applications in nanoelectronics~\cite{H.Liu13,X.Wu12,Y.Liu13,Z.Zhang15}.
 \par
However, theoretical calculations show that due to the imaginary frequencies in phononic dispersion of free-standing 2D borophene is unstable against long-wavelength periodic vibration~\cite{Xu16,Jena2017}, needing a substrate to be stabilized. A feasible proposal to dynamically stable borophene is the chemical fictionalization using surface hydrogenation. First-principles calculations show that the hydrogenation of borophene, is a viable method to stabilize borophene in the vacuum without a substrate~\cite{Xu16}. Compared with that of borophene, borophane, a monolayer of fully hydrogenated borophene has a remarkable Fermi velocity which is nearly four times higher than that of graphene \cite{Xu16}. It displays a huge electrical and magnetic anisotropy~\cite{Mzarebrop1,Mannix15,Padilhapccp16}, along with highly anisotropic mechanical properties~\cite{HWang16}.
In contrast to the buckled direction, that shows off a semiconductor behavior, along the valley direction, borophane shows a metallic trend with a linear current-voltage characteristic~\cite{Anpccp2018}.
 \par
Lower crystal symmetry of borophane, in contrast to graphene, causes the asymmetric velocity parameters which results in the two tilted Dirac cones at  ${\bf K}_D$ and  $-{\bf K}_D$, in the effective low-energy Hamiltonian of borophane.
The Bravais lattice constants of the conventional orthorhombic unit cell of the buckled structures of borophene and borophane are ($a_x = 1.62$, $a_y = 2.85$)\AA~ and ($a_x = 1.92$, $a_y = 2.81$) \AA~, respectively~\cite{Piazza14,Xu16,Tang2007} and contains 4 atoms per unit cell (see Fig.~\ref{NSshem1} (a)). Notice that the buckling height of $h= 0.96$ \AA~ in borophene reduces to $h= 0.81$ \AA, upon hydrogen adsorption in borophane~\cite{Padilhapccp16}.

\par
Due to the hybridized states of the $\sigma$ and $\pi$ bonds, 2D boron structure may be a pure single-element intrinsic superconducting material with the highest $T_c$ (higher than the liquid hydrogen temperature) on conditions without high pressure and external strain which can be modified by strain and doping~\cite{Penev-NL16,M.Gao,RC.Xiao}.

\par
A first-principles study reveals that borophene is the first known materials with high-frequency plasmons in the visible spectrum~\cite{Sadrzadeh-NL12}.
Furthermore, in this borophene polymorph, the anisotropic plasmon mode remains undamped for higher energies along the mirror symmetry direction in which the anisotropic Friedel oscillation behaves like $r^{-3}$ in the large-$r$ limit~\cite{Sadhukhan2017}.

\par
Motivated by the great interest in search of tilted Dirac material, we wish to examine whether the tilt leads to qualitatively different physics in normal-metal–superconductor hybrid junction of borophane. However, to our best knowledge, answers to these question is still lacking.
In this paper, using the extended Blonder-Tinkham-Klapwijk formalism, we study the charge and heat transport in a NS hybrid contact based on the fully hydrogenated borophene (borophane) and in the 8-$ Pmmn $ 2D boron Polymorph.
We find the dependence of the Andreev conductance on the Fermi energy and bias voltage, enable us selecting the retro configuration or specular configuration in types of Andreev reflection processes.
We numerically show that, independent of the Fermi energy, the temperature dependence of the thermal conductance in borophane can be modelled as an inverse Gaussian function, reflecting the d-wave symmetry of the borophane superconductor.
We propose a scheme for achieving negative differential thermal conductance in borophane NS junction, as a key building block of thermal circuits.
Our findings will have potential applications for transport and energy control in superconducting hybrid structures and thermal nanoscale devices.

The rest of the paper is organized as follows. In Sec.~\ref{sec:model1}, the low energy model Hamiltonian of borophane is introduced and then the method which is used to calculate the differential conductance of charge and heat transport in normal-metal–superconductor (N/S) hybrid junction of borophane is explained using the extended Blonder-Tinkham-Klapwijk formalism in Sec.~\ref{sec:model}.
In Sec.~\ref{Numr_res}, we present and describe our numerical results. Finally, we conclude and summarize our main results in Sec.~\ref{sec:concl}.

\section{Model Hamiltonian of superconducting borophene}\label{sec:model1}

\begin{figure}[t]
\begin{center}
\includegraphics[width=3.2in]{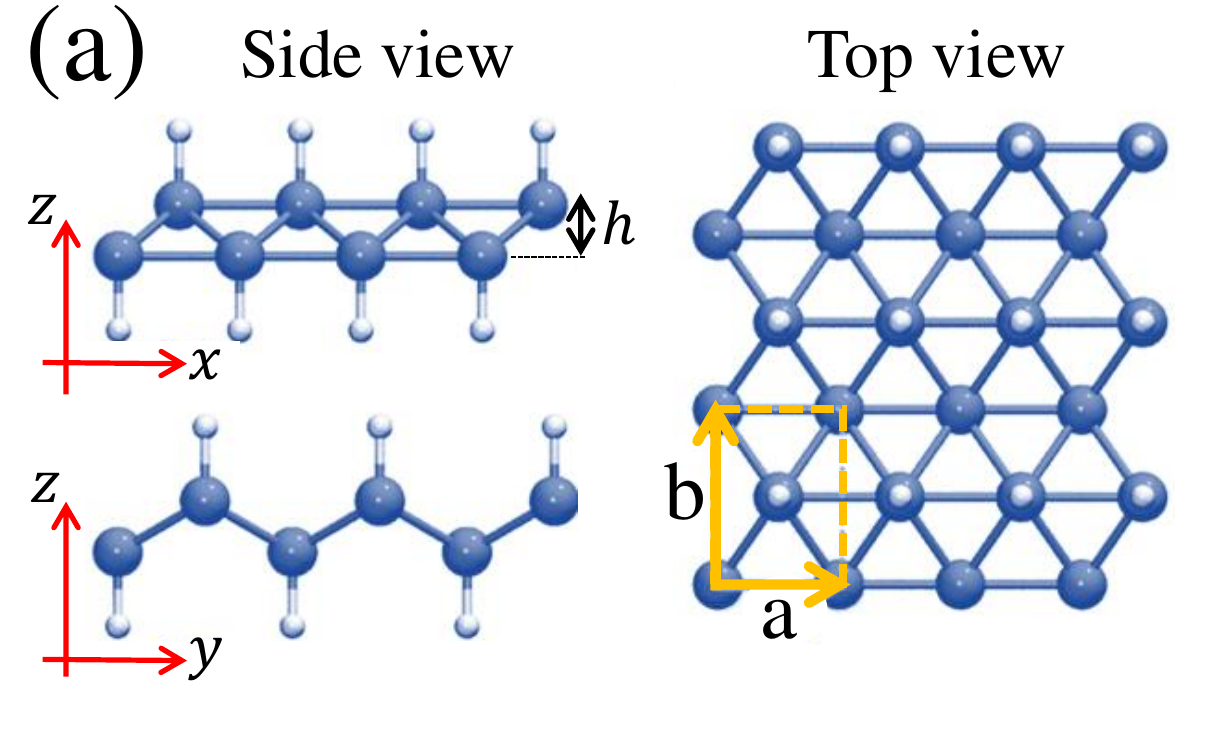}\label{NSshem1}
\includegraphics[width=3.2in]{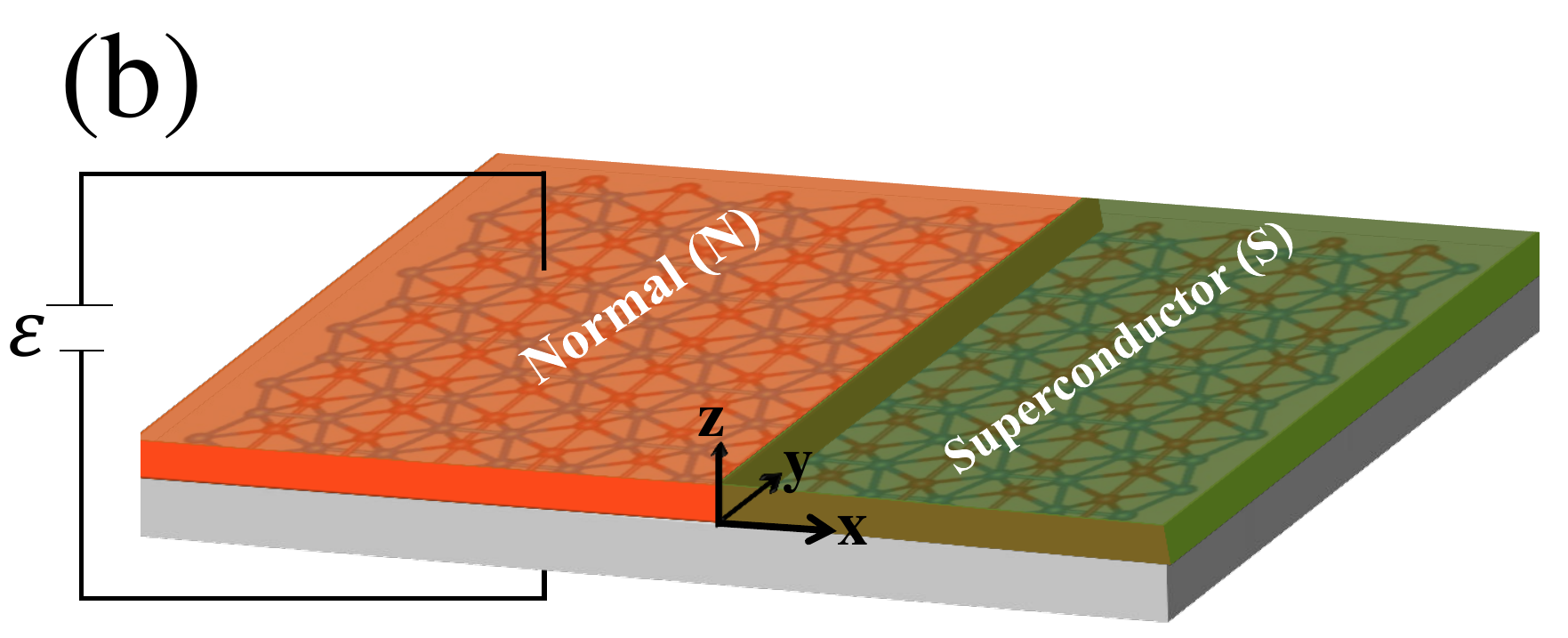}

\end{center}
\caption{ (a) Schematic illustration of the optimized ground-state lattice structure of borophane, with top view (right panel) and side views (left panel). The unit cell is indicated by the yellow-dashed rectangle, which contains two boron (B) atoms and two hydrogen (H) atoms and the basic vectors of the primitive unit cell are indicated by the yellow arrows. The blue and white balls represent B and H atoms, respectively. The buckling height is denoted by h. (b) An effective NS junction of monolayer borophane, where superconductivity is induced via a proximate superconducting lead.}
\end{figure}

In the following we consider a two dimensional normal/superconducting borophane junction, occupying the $xy$ plane while the superconducting region occupies $x>0$ and the normal region extending $x<0$. The proposed setup is schematically shown in Fig.~\ref{NSshem1} (b).

Consider the Bardeen–Cooper–Schrieffer (BCS) pairing in the S region, the electron and hole excitations are described by the Bogoliubov-de Gennes (BdG) equation which has the form \cite{BdG}

\begin{equation}
\label{DBdG1}
   \left[\begin{array}{cc}
         {\text H}_0^{\eta}+{\text V(\bm r)} & \Delta_S(\bm{r}) \\
         \Delta_{S}^{\ast}(\bm{r}) & -\mathcal T^{-1} {\text H}_0^{\eta} \mathcal T-{\text V(\bm r)}
       \end{array}
  \right]
  \left(
\begin{array}{c}
u\\
v
\end{array}
\right)
=\varepsilon\left(
\begin{array}{c}
u\\
v
\end{array}
\right),
\end{equation}

 in which $V(\bm r)=U(\bm r)-E_F$, where $E_F$ and $\mathcal{T}$ are Fermi energy and time-reversal symmetry, respectively and ${\text H}_0^{\eta}(\bm{k})$ is the effective single-particle low-energy Hamiltonian for borophane, for excitations near the Dirac points $K_\eta $; ~\cite{zabolotskiy2016strain,Sadhukhan2017,Ezawa2017}

\begin{equation}
{\text H}_0^{\eta}(\bm{k})= \eta[\hbar v_{0x} k_{x}\sigma_{x} + \hbar v_{0y}k_{y} \sigma_{y} + \hbar v_{t}k_{x} \sigma_{0}],
\label{boeqn}
\end{equation}

Here, $\sigma_{x},\sigma_{y} $ are the Pauli matrices for the pseudospin representing the lattice degree of freedom while $ \sigma_{0} $ is the $ 2 \times 2 $ identity matrix and $\eta=\pm1 $ corresponds to the ${\bf K}_D=(0.64,0)$\AA$^{-1}$ and $-{\bf K}_D=(-0.64,0)$\AA$^{-1}$ valley.
The suggested values of the direction-dependent velocities, in units of $\left(\times 10^{5}\,m/s \right)$, are specified as $ v_{0x} = 19.58 $, $ v_{0y} = 6.32 $, and $ v_{t} = -5.06 $.

The superconducting order parameter $\Delta_S(\bm{r})$ couples time-reversed electron and hole states, which can be expressed as
\begin{eqnarray}\label{eq:array}
\Delta_S(\bm{r}) =\left\{\begin{array}{ll}
0&\text{if $x<0$, }\\
\Delta_0 &\text{if $x>0$}\\
\end{array}\right.
\end{eqnarray}

and since the zero of potential is arbitrary, we assume the electrostatic potential $U(\bm r)$ as follows

\begin{eqnarray}\label{eq:array}
U(\bm r)=\left\{\begin{array}{ll}
0&\text{if $x<0$, }\\
U_0&\text{if $x>0$}\\
\end{array}\right.
\end{eqnarray}

The time-reversal symmetry requires $\mathcal TH_0^{\eta}(\bm k)\mathcal T^{-1}=H_0^{-\eta}(-\bm k)$, thus in the absence of a magnetic field, the time-reversal operator interchanges the valleys ${\bf K}_D$ with $-{\bf K}_D$.
Neglecting the intervalley scattering and due to spin-valley degeneracy it suffices to consider one of these two sets $H_0^{\eta}$ or $H_0^{-\eta}$. Therefore, the (8$\times$8)-matrix BdG equation \ref{DBdG1} reduces to two decoupled sets of four-dimensional Dirac-Bogoliubov-De Gennes (DBdG) equation of the form

\begin{equation}
\label{DBdG2}
   \left[\begin{array}{cc}
         {\text H}_0^{\pm}+{\text V(\bm r)} & \Delta_S(\bm{r}) \\
         \Delta_{S}^{\ast}(\bm{r}) & -{\text H}_0^{\pm}-{\text V(\bm r)}
       \end{array}
  \right]
  \left(
\begin{array}{c}
u\\
v
\end{array}
\right)
=\varepsilon\left(
\begin{array}{c}
u\\
v
\end{array}
\right),
\end{equation}

The energy dispersion for quasiparticles in the superconducting region is written as

\begin{eqnarray}
&&E_{S}^{h(e)}(\emph{\textbf{k}})= \tau \sqrt{\Delta_0^{2}+[E_F-U_0-v_{t}k_{x} \pm \sqrt{v_{0x}^{2}k_{x}^{2}+v_{0y}^{2}k_{y}^{2}}]^2},\nonumber \\
\end{eqnarray}

in which $\tau=$1(-1), denotes the conduction (valence) band in borophane and the superscripts "e" and "h" denote the electronlike and holelike excitations, respectively.
%%%%%%%%%%%%%%%%%%%%%%%%%%%%%%%%%%%%%%%%%%%%%%%%%%%%%
The energy dispersion for ${\bf K}_D$-valley is shown in Fig.~(\ref{NSshem1}), which is tilted along $k_x$ due to the presence of $v_t$ term.
In $-{\bf K}_D$-valley, the tilting is in opposite direction and in contrast to graphene, Dirac cones are
anisotropic. In particular, we note that the particle-hole symmetry is broken in borophane due to the tilted feature of the Dirac cones, because
$ {\mathcal P} {\text H}_0^{\eta}(k) {\mathcal P}\neq -{\text H}_0^{\eta}(-k)$, where $\mathcal{P}=\sigma_{0}\tau_x \mathcal{K}$ and $\mathcal{K}$ is the complex conjugate operator.

In the normal borophane region, superconducting order $\Delta_0$ vanishes and there are two electron modes and two hole modes with energy dispersions given as

\begin{eqnarray}\label{eq:diseh}
E_{c}^{e(h)}(\emph{\textbf{k}})&=& \hbar v_{t}k_{x}+ \hbar\sqrt{v_{0x}^{2}k_{x}^{2} + v_{0y}^{2}k_{y}^{2}}\mp E_F,\\
\label{eq:diseh2}
E_{v}^{e(h)}(\emph{\textbf{k}})&=& \hbar v_{t}k_{x}- \hbar\sqrt{v_{0x}^{2}k_{x}^{2} + v_{0y}^{2}k_{y}^{2}}\mp E_F,
\end{eqnarray}
in which $c(v)$ denote the conduction (valence) band in borophane and $k=\sqrt{k_x^2+k_y^2}$. The anisotropic and tilted Dirac crossing along the $ \Gamma $-X direction in the rectangular Brillouin zone of borophane is obtained from the band dispersions Eqn.\ref{eq:diseh} and Eqn.\ref{eq:diseh2}, as

\begin{figure*}\label{fig:BTK-scattering}
  \includegraphics[width=0.7\linewidth]{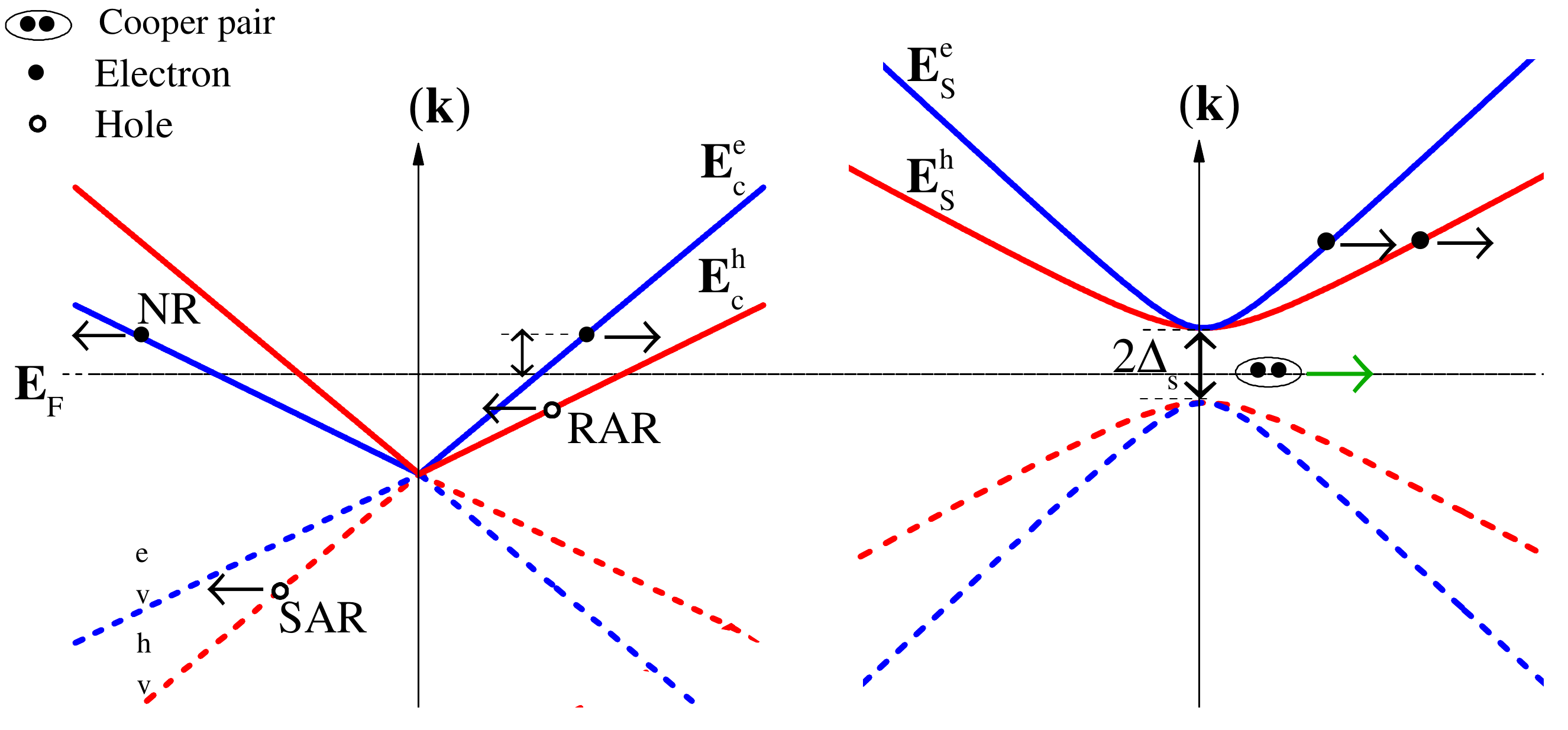}
\caption{Schematic illustration of the reflection and transmission processes in the NS junction of borophane. The filled and open circles denote the electron and a hole, respectively. When the excitation energy becomes smaller than the superconducting order parameter ($\varepsilon<\Delta_0$) there are no propagating modes in the superconducting side, so that an electron injected from the normal region is reflected back either as an electron or a hole. However, there are two scattering process for holes: if the excitation energy $\varepsilon$ becomes smaller (larger) than the normal region Fermi energy, $\varepsilon<E_F$ ($\varepsilon>E_F$), retroreflection-RAR (specular reflection-SAR) occurs. When Andreev reflection takes place, the transmitted Cooper pair is located at the Fermi level of the superconductor.
In the S region (right panel), the green arrow represents an evanescent mode (when $\varepsilon<\Delta_0$ ), while the black ones denote the propagating modes (when $\varepsilon>\Delta_0$ ).}
\end{figure*}

The eigenfunctions of the Dirac-Bogoliubov quasiparticles with energy $\varepsilon$ are given by

\begin{eqnarray}
\psi_{\tau }^{S+}=\ e^{ ik_{s}^{e} x} e^{iqy}
\left(
\begin{array}{c}
 u_{+}  e^{- i\beta_{{k}_s}^e} \\
      u_{+}  \\
  v_{-} e^{- i\beta_{{k}_s}^e} e^{- i\phi} \\
   v_{-}  e^{- i\phi}
\end{array}
\right),
\\\nonumber\\
\psi_{\tau }^{S-}=\ e^{-i k_{s}^{h} x} e^{iqy}
\left(
\begin{array}{c}
v_{-}  e^{- i\beta_{{k}_s}^h} \\
   v_{-}    \\
  u_{+}  e^{- i\beta_{{k}_s}^h} e^{- i\phi} \\
   u_{+}  e^{- i\phi}
\end{array}
\right),
\end{eqnarray}

from the $E_S^e$ and $E_S^h$ branches of the spectrum (see Fig.\ref{fig:BTK-scattering}, right panel), respectively, which describing right-moving electron and left-moving holelike quasiparticles that either decay exponentially as $x\rightarrow \infty$ (for subgap solutions when $\varepsilon<\Delta_0$) or propagate along the $x$ direction (for supragap solutions when $\varepsilon>\Delta_0$). The coherence factors $ u_{+}, v_{-}$ are generically written as \cite{sudbo}

\begin{eqnarray}
u_{+} = \sqrt{\frac{1}{2}\Big(1 + \frac{\sqrt{E^2-|\Delta_S|^2}}{E}\Big)}, \\
v_{-} = \sqrt{\frac{1}{2}\Big(1 - \frac{\sqrt{E^2-|\Delta_S|^2}}{E}\Big)}.
\end{eqnarray}

The longitudinal wave vector ${k}_s^{e(h)}$ for electronlike (holelike) qausiparticles in the S region are the solutions of the energy-momentum relation, which can be obtained by solving the following equation

\begin{eqnarray}
a_4  x^4 + a_3  x^3 + a_2 x^2 + a_1 x + a_0=0
\end{eqnarray}

in which

\begin{eqnarray}
a_0 &=& b_1^2-2 b_1 \varepsilon^2+\varepsilon^4+4 b_2^2 v_t^2-4v_{0y}^2 v_t^2 q^2,\nonumber\\
a_1 &=& 4 b_2 v_t (-b_1 + \varepsilon^2 + 2 v_{0y}^2 v_t^2),\nonumber\\
a_2 &=& 2 b_1 b_3 - 2 \varepsilon^2 b_3 + 4 b_2^2 (-v_{0x}^2+ v_t^2) - 4 v_{0y}^2 v_t^2 q^2,\nonumber\\
a_3 &=& 4 b_2 v_t( v_{0x}^2 - b_3 ),\nonumber\\
a_4 &=&b_3^2-4v_{0x}^2 v_t^2,
\end{eqnarray}

Here, $b_1 = \Delta_S^2 + b_2^2 + v_{0y}^2 q^2$, $b_2=E'_F - Ur$ and $b_3= v_{0x}^2+ v_t^2$.
Inside the N region, the solutions of BdG equation are two states of the form
\begin{equation}
\label{psie}
\psi^{e\pm}=\frac{1}{\sqrt{N_e}}\ e^{\pm ik_x^{e} x} e^{iqy}
\left(
\begin{array}{c}
\tau_e  ~e^{\mp i\beta_{{k}}^e}\\
1\\
0\\
0
\end{array}
\right),
\end{equation}
for the conduction band electrons and
\begin{equation}
\label{psih}
\psi^{h\pm}=\frac{1}{\sqrt{N_h}}\ e^{\mp  i \alpha k_x^{h} x} e^{iqy}
\left(
\begin{array}{c}
0\\
0\\
 \tau_h ~e^{\pm i \alpha \beta_{{k}}^h}\\
1
\end{array}
\right),
\end{equation}

for the conduction ($\tau_h=1$) or valence band ($\tau_h=-1$) holes, depending on the reflection type. Here, $\beta_{k}^{e(h)}=\tan^{-1}\left[ v_{0y} q/v_{0x} k_x^{e(h)} \right]$ with $\alpha=1$ for retroreflection (conduction band holes) and $\alpha=-1$ for specular reflection (valence band holes),
where $k_x^e$ and $k_x^h$ are corresponding to the x-component of wave vectors.
It should be noted that, due to the translational invariance in the y direction, the corresponding component of momentum, $q$ is conserved. The prefactors $1/{\sqrt{N_{e}}}, 1/{\sqrt{N_{h}}}$ ensure that all states carry the same amount of quasiparticle current. In order to obtain these normalizing factors, we calculate the expectation value of carrier velocity using the formula
$j_x=\psi^\ast \frac{\partial H}{\hbar\partial {k_x}} \psi=v_x$. Furthermore, the $x$ and $y$ components of the velocity operator is defined as ${\bf{v}} = \frac{1}{\hbar}\nabla_{\bf{k}} E(\bf{k})$ and can be derived as

\begin{eqnarray}
v_{x,\tau}^{e(h)}&=&v_{t}+\tau \frac{v_{0x}^{2}k_x^{e(h)}}{\sqrt{v_{0x}^{2}{k_x^{e(h)}}^2+v_{0y}^{2}q^2}},\\
v_{y,\tau}^{e(h)}&=&\tau \frac{v_{0y}^{2}q}{\sqrt{v_{0x}^{2}{k_x^{e(h)}}^2+v_{0y}^{2}q^2}},
\end{eqnarray}

\section{MODEL AND THEORY}\label{sec:model}

Let us consider an incident electron from the normal side of the junction with energy $\varepsilon$ and transverse momenta q. Taking into account both Andreev and normal reflection processes, the wave functions in the normal and superconducting regions, can be written as
\begin{eqnarray}\label{contin}
\psi_{N}&=&\psi^{e+}_{N}+r_{e}\ \psi^{e-}_{N}+r_{A}\ \psi^{h-}_{N},\\
\psi_{S}&=&t_e\ \psi^{S+}+t_h\ \psi^{S-},
\end{eqnarray}

Here, $\psi^{e(h)\pm}_{N}$ and $\psi^{S\pm}$ are the solutions of BdG equation for the quasiparticles inside N and S regions, respectively.
$r_{e}$ and $r_{A}$ denote the reflection coefficients of normal and Andreev reflections, respectively.

The boundary conditions for the wave functions and current conservation in the x direction, at the interface can be
written as
\begin{eqnarray}\label{bc1}
\psi_N |_{x=0} &=&\psi_S |_{x=0},\nonumber\\
  v_N^x \psi_N |_{x=0} &=& v_S^x \psi_S |_{x=0}
\end{eqnarray}

which leads to the analytical expressions for the reflection coefficients $r_e$ and $r_A$ as:

\begin{widetext}
\begin{eqnarray}\label{Re-h}
r_e &=& \frac{e^{-i \beta_{k}^e}[-U_+^2[e^{i( \alpha \beta_{k}^h+ \beta_{k_s}^e)}+\tau_h e^{i( \beta_{k}^e+ \beta_{k_s}^h)}]+V_-^2 [e^{i( \alpha \beta_{k}^h+ \beta_{k_s}^h)}+\tau_h e^{i( \beta_{k}^e+ \beta_{k_s}^e)}]+(U_+^2-V_-^2)[e^{i( \beta_{k}^e+ \alpha \beta_{k}^h)}+\tau_h e^{i( \beta_{k_s}^h+ \beta_{k_s}^e)}]]}{U_+^2[-e^{i( \alpha \beta_{k}^h+ \beta_{k'}^e +\beta_{k_s}^e)}+\tau_h e^{i \beta_{k_s}^h}]+V_-^2[e^{i( \alpha \beta_{k}^h+ \beta_{k'}^e +\beta_{k_s}^h)}-\tau_h e^{i \beta_{k_s}^e}]+(U_+^2-V_-^2)[\tau_h e^{i( \beta_{k'}^e+ \beta_{k_s}^h+\beta_{k_s}^e)}- e^{i\alpha \beta_{k}^h }]},\nonumber\\
r_h&=& \frac{U_+ V_- e^{-i(\phi+\beta_{k}^e+ \alpha \beta_{k}^h)}(1+e^{i(\beta_{k}^e+\beta_{k'}^e)})(e^{i \beta_{k_s}^h}-e^{i \beta_{k_s}^e})}{U_+^2[-e^{i( \alpha \beta_{k}^h+ \beta_{k'}^e +\beta_{k_s}^e)}+\tau_h e^{i \beta_{k_s}^h}]+V_-^2[e^{i( \alpha \beta_{k}^h+ \beta_{k'}^e +\beta_{k_s}^h)}-\tau_h e^{i \beta_{k_s}^e}]+(U_+^2-V_-^2)[\tau_h e^{i( \beta_{k'}^e+ \beta_{k_s}^h+\beta_{k_s}^e)}- e^{i\alpha \beta_{k}^h }]},\nonumber\\
\end{eqnarray}
\end{widetext}

\subsection{Andreev conductance}\label{sec:AR-G1}

In this section, we investigate properties of the differential tunneling conductance of the borophane NS junction using the well-known Blonder-Tinkham-Klapwijk (BTK) formula \cite{double,BTK}

\begin{eqnarray}\label{GAR}
G(\varepsilon)=G_{0}\sum_{{\pm \bf K}_D}\sum_{q}\left[1 - \left| {r(\varepsilon,q )} \right|^2  +
\frac{|p_\text{h}|\cos\phi_\text{A}}{|p_\text{e}|\cos\phi} \left| {r_A (\varepsilon,q )} \right|^2 \right],\nonumber\\
\end{eqnarray}

where $G_0=\frac{4e^2}{h}{N}(eV)$ is the ballistic conductance of a monolayer borophane with ${N}$ transverse modes [\onlinecite{AR-G1}, \onlinecite{Jiang2008}], and $p_\text{e(h)}$ is the wavevector of the electronlike (holelike) quasiparticles, inside the N region. The first summation is over two valleys in the band structure of borophane and the second one is over all transverse modes of the NS junction.

The density of states  $N(\varepsilon^{\tau})$, can be obtained by solving the following equation

\begin{eqnarray}\label{eq-54}
 {N}(\varepsilon) &=& \displaystyle \frac{1}{(2\pi)^2} \int_0^\infty k'\, dk'
  \delta(\varepsilon-\varepsilon_{\vec{k}'})
\end{eqnarray}

Performing this integral over energy, one finds the following expansion for the density of states of borophane:

\begin{eqnarray}\label{eq-54}
  N(\varepsilon^{\tau}) = \frac{k^{\tau}(\varepsilon,\phi) }{v_t\cos(\phi) + \sqrt{v_{0x}^2 \cos^2(\phi)+ v_{0y}^2 \sin^2(\phi)}},
\end{eqnarray}

where $\phi=\tan^{-1}(k_y/k_x)$ and the wave vector $ k^{\tau}(\varepsilon,\phi)$ is given by

\begin{eqnarray}\label{eq-54}
 k^{\tau}(\varepsilon,\phi) =\varepsilon  \frac{-2 v_t\cos(\phi) +\tau \sqrt{2}\sqrt{ v_x^2 +  v_y^2 +  (v_x^2-  v_y^2) \cos(2\phi)}}{v_x^2 + v_y^2 - v_t^2 + (v_x^2 - v_y^2 - v_t^2) \cos(2\phi)},\nonumber\\
\end{eqnarray}

\subsection{Heat transport by normal/superconducting borophane junctions}\label{sec:AR-G1}
In order to investigate the thermal transport properties of the proposed N/S structure, assuming a temperature gradient $\Delta T$ through the junction, using relation $\kappa=\lim_{\Delta T\to 0}J_{Q}/\Delta T$, with $J_{Q}$ the heat current density, we present the behavior of the thermal conductance given by~\cite{Bardas95,Yokoyama08}

\begin{widetext}
\begin{eqnarray}\label{kapa}
\kappa  = \kappa_0\sum_{{\pm \bf K}_D}\sum_{q} \int_{ 0 }^\infty d\varepsilon { { \left[1 - \left| {r(\varepsilon,q )} \right|^2  -
\frac{|p_\text{h}|\cos\phi_\text{A}}{|p_\text{e}|\cos\phi} \left| {r_A (\varepsilon,q )} \right|^2 \right]\frac{{\varepsilon^2 }}{{(k_B T)^2 \cosh ^2 (\frac{\varepsilon}{{2k_B T}})}}} },
\end{eqnarray}
\end{widetext}
in which $ \kappa_0=\frac{k_B W}{8\pi^2 \hbar}$ is a constant parameter corresponding to the N/N thermal conductance of a sheet of monolayer borophane of width W \cite{kashiwaya96}.

We replace the zero-temperature superconducting order parameter $\Delta_S$ in Eq. (\ref{DBdG2}) with the temperature-dependent energy-gap function of $\Delta_S(T)=1.76 \ k_B T_C\ \tanh{(1.74\sqrt{{T_C}/{T}-1})}$, where $T_C$ is the critical temperature of the superconductor.

\section{Numerical results}\label{Numr_res}

In this section, based on Eqs.~\ref{GAR} and \ref{kapa} we present our numerical results for the AR process of the NS hybrid structure of borophane in the physical regime. Since attaining the regime $E_{F}\ll\Delta_{0}$, in experiments may be difficult, so it is of importance to consider the regime of comparable $E_{F}$ and $\Delta_{0}$, in which retro-reflection changes to specular Andreev reflection. We set $\Delta_0=0.01$eV for zero temperature order parametere in all our results presented in this section, except in the case of thermal conductance, with $E_{F}$ and $E'_{F}$, the Fermi energies in the normal and superconducting regions, respectively.
Let us now consider the regime where the Fermi surfaces of the normal metal and the superconductor is aligned $U(\bm{r})=0$.

\begin{figure}[]
\begin{center}
\includegraphics[width=3.5in]{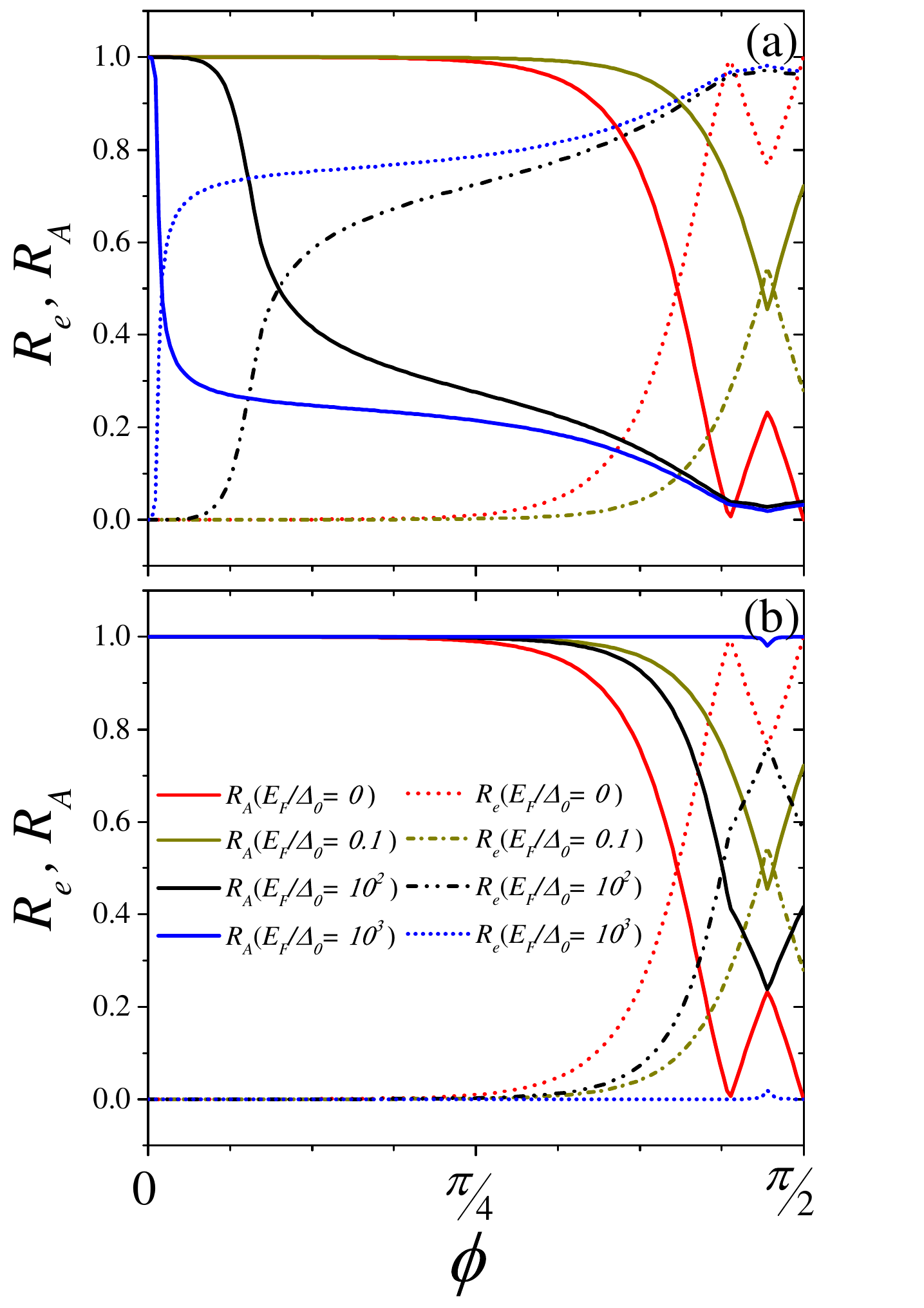}
\end{center}
\caption{\label{ReA}
Normal and Andreev reflection probabilities versus the angle of incidence for $\varepsilon/\Delta_{0}=0.01$, for several values of the normal region Fermi energy ($E_F/\Delta_{0}$). (a) for $E'_{F}/\Delta_{0}=10$ and (b) for $E'_{F}/\Delta_{0}=10^3$}
\end{figure}

To calculate the probability of the electron-hole conversion for subgap energies ($\varepsilon < \Delta_{0}$), we show the behavior of the probability of normal and Andreev process for an incident electron with a subgap energy $\varepsilon/\Delta_{0}=0.01$, in terms of the angle of incidence, in Fig.~\ref{ReA} for several values of the normal region Fermi energy $E_F$. Panel (a) is for $E'_{F}/\Delta_{0}=10$ and panel (b) is for $E'_{F}/\Delta_{0}=10^3$.

Since the transmission into the superconductor region is forbidden for subgap energies ($\varepsilon<\Delta_{0}$), one obviously verifies that $|r|^{2}+|r_{A}|^{2}=1$. The electron-hole conversion at normal incidence ($\phi=0$), happens with unit probability ($|r_{A}|^{2}=1$ ), regardless of the amount of the Fermi energy of the normal borophane. This is complectly different from usual normal-metal-superconductor contact, in which Andreev reflection is suppressed at any angle of incidence, if the Fermi wave lengths at the two sides of the interface are very different.

More interestingly, in the heavily doped superconducting regime ($E'_{F}/\Delta_{0}\gg1$), for sufficiently large absolute values of $E_F$ ($E_{F}/\Delta_{0}\gg1$), only normal Andreev reflection is present for any incident angle.

\begin{figure}[]
\begin{center}
\includegraphics[width=3.7in]{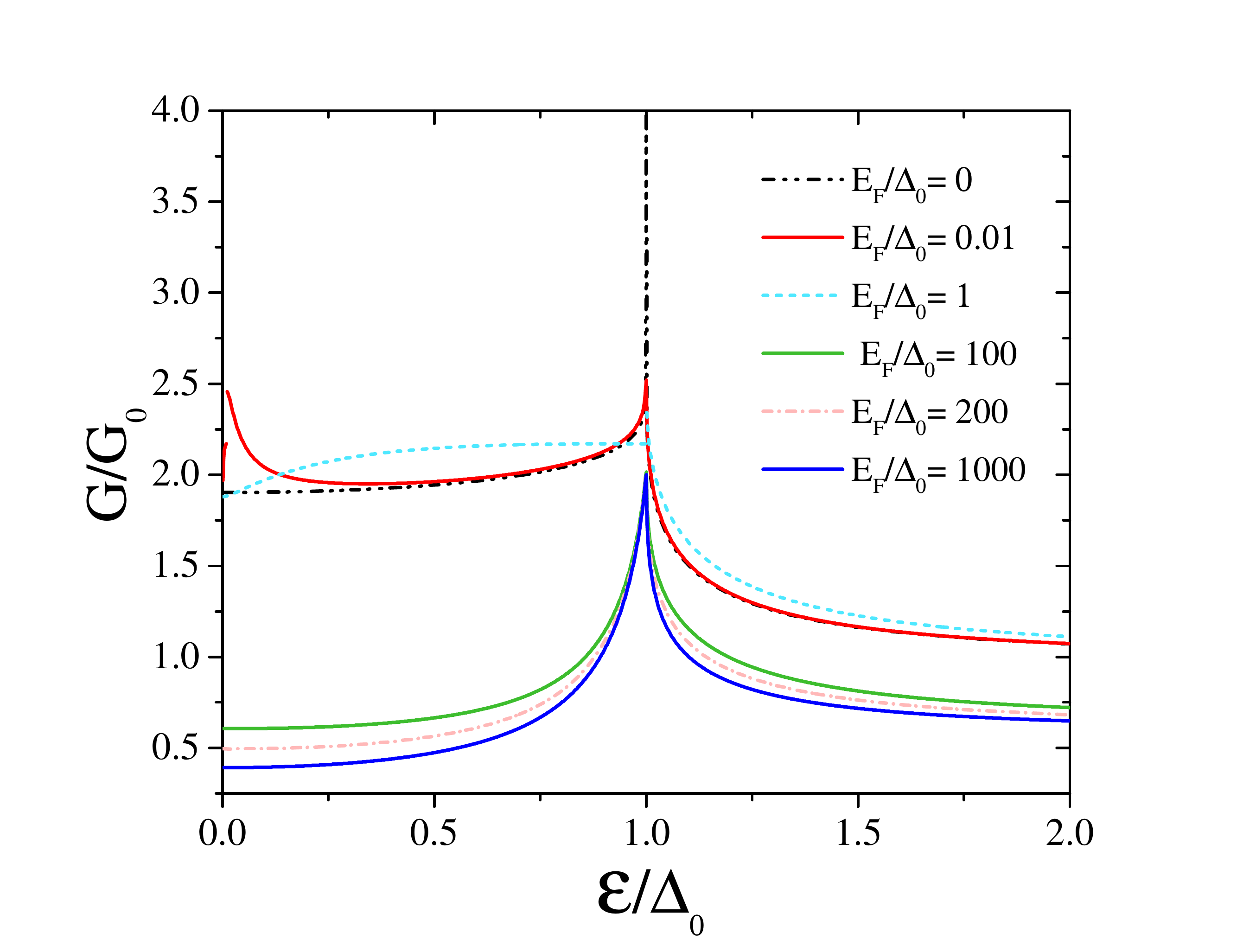}
\end{center}
\caption{\label{GeV1}
Normalized  Differential conductance (in units of the ballistic value $G_{0}=4Ne^2/h$) of the borophane NS junction, for several values of the normal region Fermi energy $E_F/\Delta_{0}$) for $E'_{F}/\Delta_{0}=10$.}
\end{figure}

As the experimentally measurable Andreev conductance contains both the conductance due to Andreev reflection and electron tunneling, we present the behavior of the normalized total Andreev conductance of the N/S structure $G/G_0$ as a function of the bias voltage $\varepsilon/\Delta_0$ in Fig. \ref{GeV1}, for different values of normal borophane Fermi energy $E_{F}/\Delta_{0}$.

As seen in this figure, the conductance displays two limiting behaviors for $E_{F}\gg\Delta_{0}$ or $E_{F}\ll\Delta_{0}$. At $E_{F}=0$ a sharp coherence peak is appeared in the differential conductance. For the case of $E_{F}\ll\Delta_{0}$, behavior of the differential conductance is similar to the normal metal-insulator-superconductor (NIS) junction of graphene~\cite{sengupta_prl_06}.
As usual for an NS junction \cite{Tin04}, the conductance has a singularity at $\varepsilon=\Delta_{0}$,
As bias voltage increases toward $\varepsilon/\Delta_{0}=1$, the conductance becomes larger toward $G=2G_0$ then with increasing the bias voltage, the conductance becomes lower toward a saturation conductance.
Interestingly, a zero-bias conductance peaks appears in the conductance spectra. This two peak structure is just similar to the case of normal metal—insulator d-wave superconductor junction~\cite{Tanaka95}.

\begin{figure}[]
\begin{center}
\includegraphics[width=3.7in]{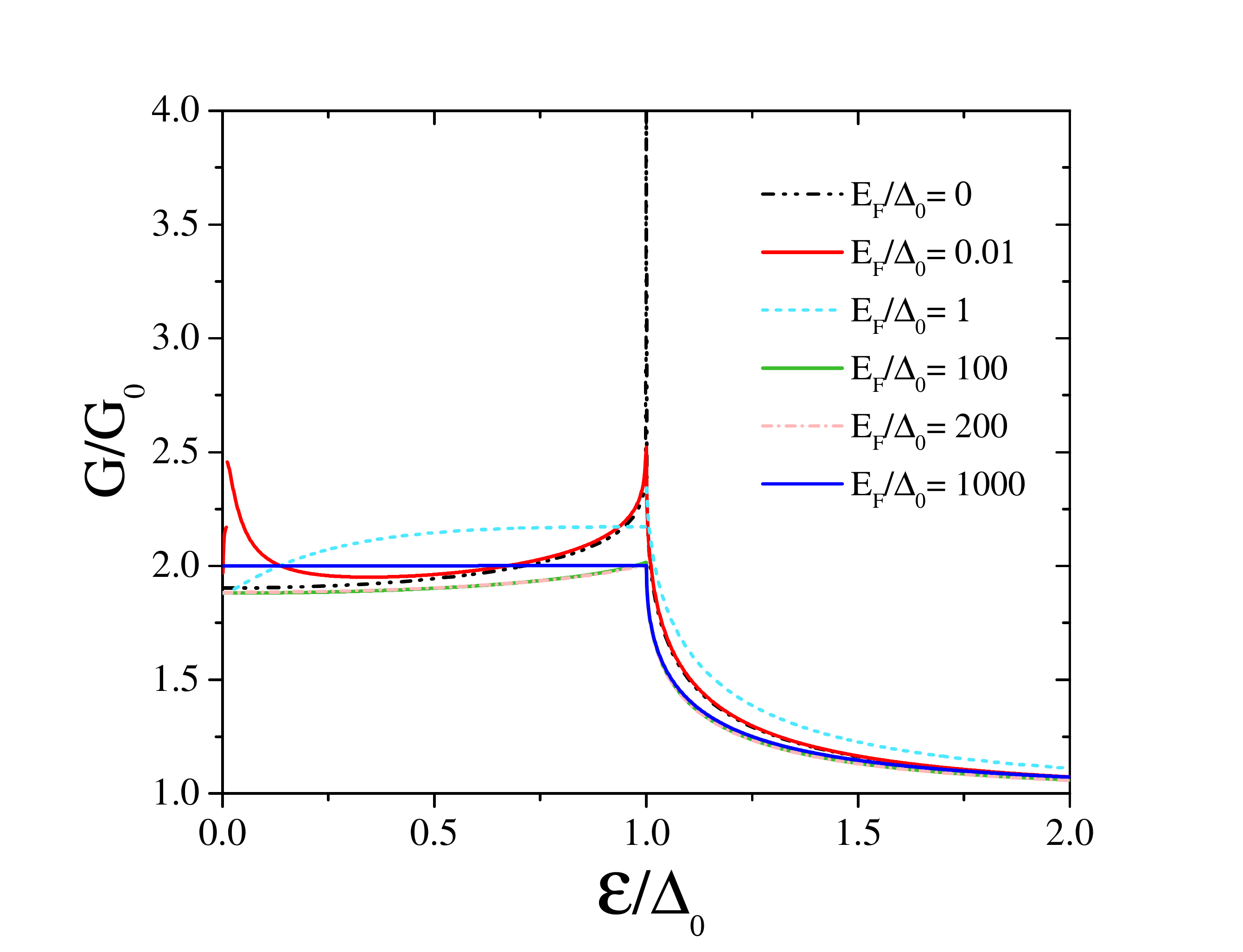}
\end{center}
\caption{\label{GeV2}
Normalized  Differential conductance (in units of the ballistic value $G_{0}=4Ne^2/h$) of the borophane NS junction, for several values of the normal region Fermi energy $E_F/\Delta_{0}$) for $E'_{F}/\Delta_{0}=10^3$.}
\end{figure}

In Fig. \ref{GeV2}, we briefly explore the behavior of the differential conductance of a NS junction of borophane, in the heavily doped superconducting regime ($E'_{F}/ \Delta_{0}\gg1$), where $E'_{F}/\Delta_{0}=1000$. In contrast to the above considered limit (low doping of the superconducting region), the conductance peak related to the bias voltage of $\varepsilon/\Delta_{0}=1$ becomes smooth as $E_{F}$ increases.
It has been shown that, for subgap energies ($\varepsilon/\Delta_{0}<1$), in the regime $E'_{F}=E_{F}$, the standard situation of perfect Andreev reflection is recovered, with a sharp drop at the gap edge, corresponding to the onset of quasiparticle transmittance into the superconductor side.
Exactly the same as unconventional anisotropic d-wave superconductor-graphene junctions, the subgap conductance is always close to $2G_0$, but becomes more constant with increasing $E_{F}$.

\begin{figure}[]
\begin{center}
\includegraphics[width=3.7in]{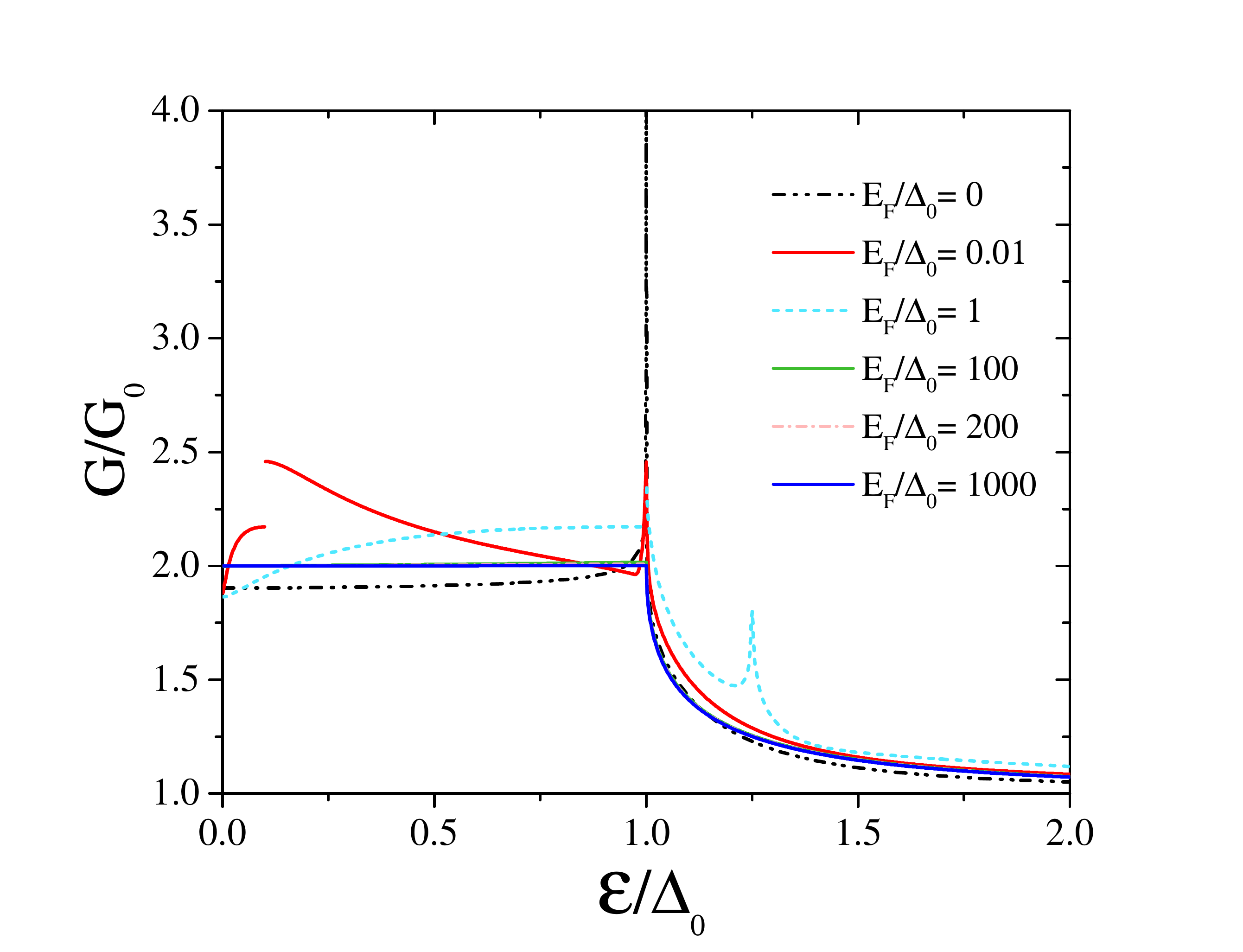}
\end{center}
\caption{\label{GeV3}
Same as Figs.~\ref{GeV1} and \ref{GeV2}, but now we have fixed $E'_{F}=E_{F}$ for all curves. }
\end{figure}

A similar qualitative behavior for the differential conductance occurs when $E'_{F}=E_{F}$. As we can see in Fig.~\ref{GeV3}, a two peak structure appears for the case where $E_{F}\ / \Delta_{0}=1$. In the limit of very large $E_{F}\ / \Delta_{0}$, the standard situation of perfect Andreev reflection is recovered.

The same as Figs.~\ref{GeV1} and \ref{GeV2}, but now for when $E'_{F}=E_{F}$, is plotted in Fig.~\ref{GeV3}. We give the results in the two regimes, $\varepsilon/\Delta_{0}<1$ and $\varepsilon/\Delta_{0}>1$. For $\varepsilon/\Delta_{0}<1$, the conductance related to the nearly zero Fermi energy ($E_{F}\rightarrow0$), increases with increasing the bias voltage $\varepsilon$, reached to value $G=4G_0$ for $\varepsilon/\Delta_{0}=1$. In the limit of $E'_{F}=E_{F}$, however, the conductance related to each bias voltage is identical.

\begin{figure}[t]
\begin{center}
\includegraphics[width=3.7in]{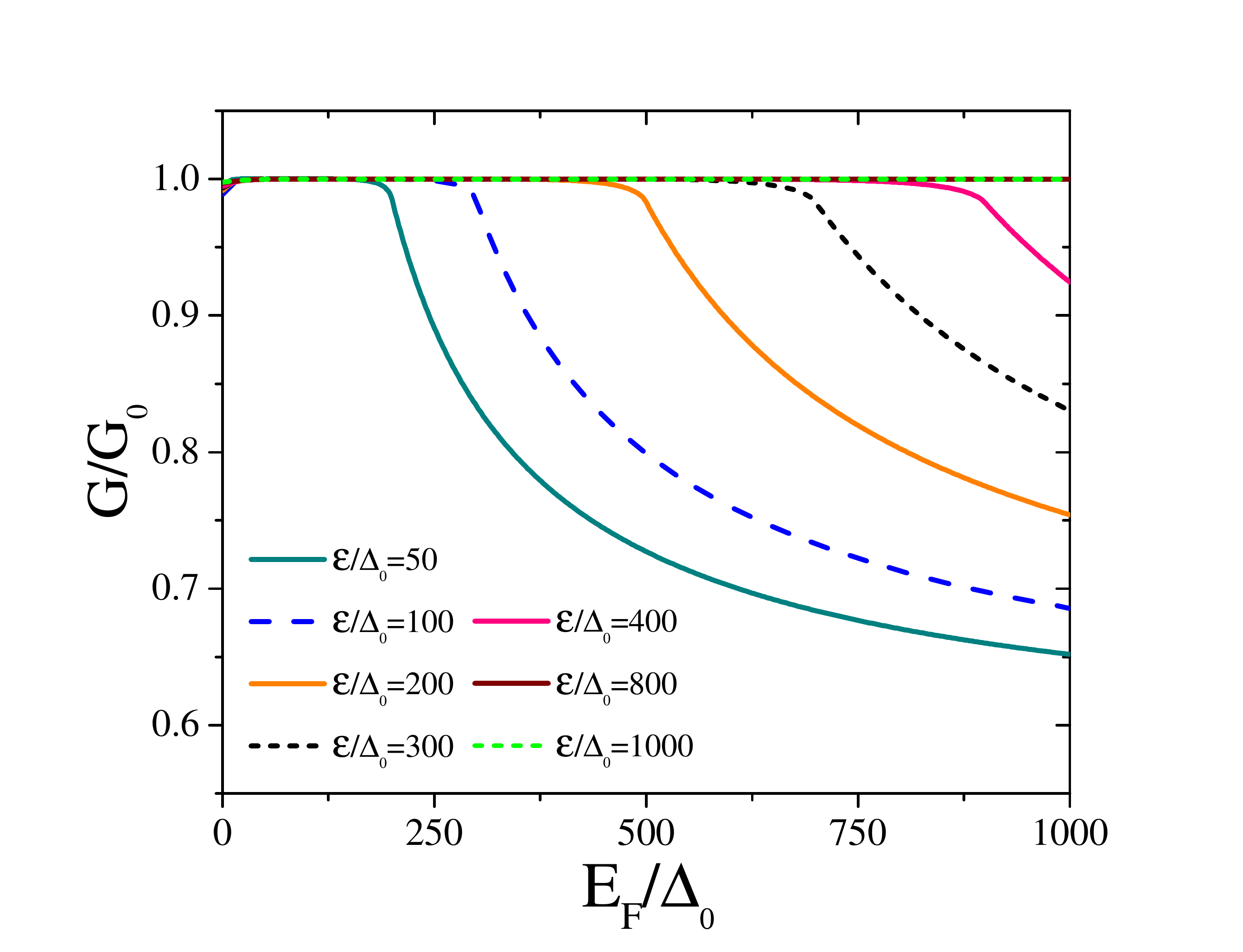}
\end{center}
\caption{\label{GEF1}
Dependence of the normalized  differential conductance (in units of the ballistic value $G_{0}=4Ne^2/h$) of the borophane NS junction, on the  Fermi energy $E_F/\Delta_0$, for $E'_{F}/\Delta_{0}=100$.}
\end{figure}

For $\varepsilon/\Delta_{0}>1$, behavior of the differential conductance is entirely different. With increasing the bias voltage, the Andreev conductance saturated to a constant value $G=G_0$, where for $E_F/\Delta_{0}>500$, the differential conductance is approximately independent of the Fermi energy of the normal borophane.

\begin{figure}[t]
\begin{center}
\includegraphics[width=3.7in]{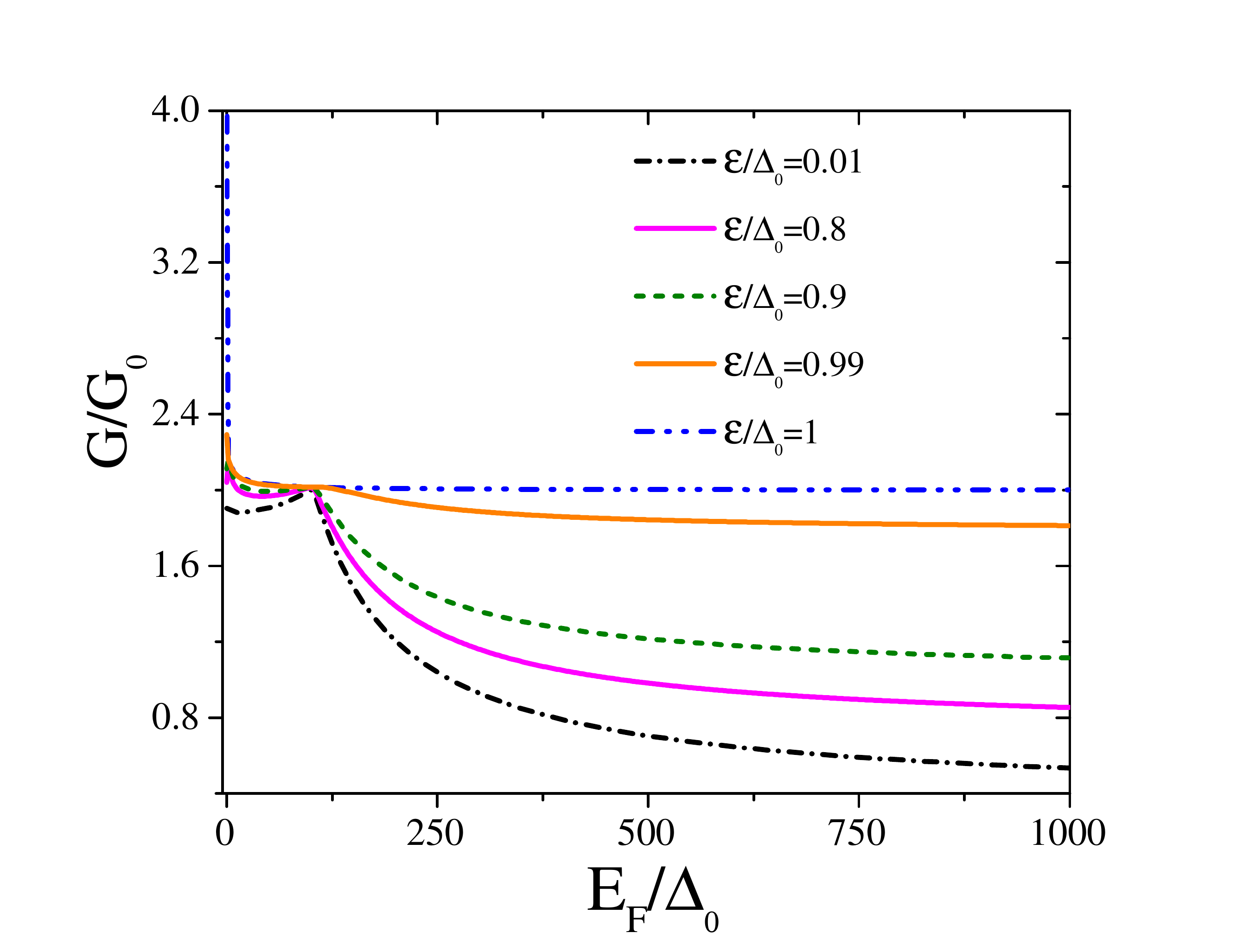}
\end{center}
\caption{\label{GEF2}
Dependence of the normalized  differential conductance (in units of the ballistic value $G_{0}=4Ne^2/h$) of the borophane NS junction, on the  Fermi energy of the normal borophane $E_F/\Delta_0$, for $E'_{F}/\Delta_{0}=100$. Insets show a zoomed view of the data within the dashed-rectangles}.
\end{figure}

On the other hand, study of the dependence of the differential Andreev conductance $G/G_0$, on the Fermi energy of the normal region ($E_{F}/\Delta_{0}$), is crucial for experimental accessibility of our proposals. We give the results for $E'_{F}/\Delta_{0}=100$, in the two regimes $\varepsilon/\Delta_{0}>1$ and $\varepsilon/\Delta_{0}<1$ in the Figs.~\ref{GEF1} and ~\ref{GEF2}, respectively.
For $\varepsilon/\Delta_{0}>1$, the conductance related to the nearly zero Fermi energy ($E_{F}\rightarrow0$), for each value of the bias voltage $\varepsilon$, reached to value $G=G_0$. In the high bias voltage regime ($\varepsilon/\Delta_{0}\gg1$), Andreev conductance is independent of the normal region Fermi energy.
For $\varepsilon/\Delta_{0}<1$, the Andreev conductance related to the zero Fermi energy ($E_{F}\rightarrow0$), has an increasing behavior with increasing the bias voltage $\varepsilon$, and reaches a maximum value ($G(E_F=0)=4G_0$) at zero Fermi energy.

In the limit of $E'_{F}=E_{F}$, however, the conductance related to each bias voltage reaches a constant value ($G(E'_{F}=E_{F})=2G_0$). Turning on the normal region Fermi energy, away from the zero Fermi energy, the Andreev conductance suddenly drop to a lower constant value for $\varepsilon/\Delta_{0}=1$. It is seen that in a certain Fermi energy, with increasing the bias voltage $\varepsilon/\Delta_{0}$, the conductance increases such that saturates toward a constant value $G=2G_0$.

\begin{figure}[t]
\begin{center}
\includegraphics[width=3.3in]{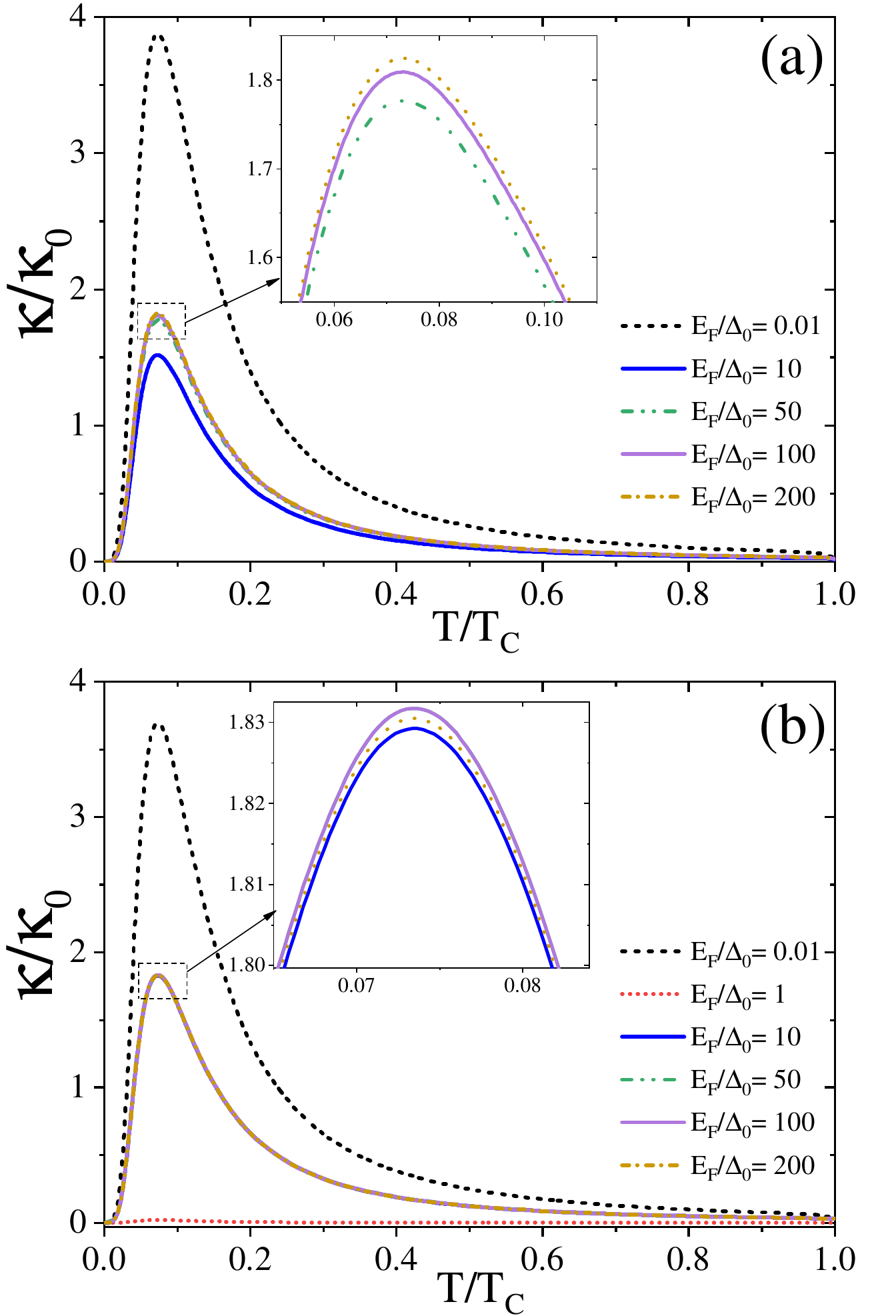}
\end{center}
\caption{\label{kth1}
Normalized thermal conductance of the borophane NS junction, as a function of the temperature, in units of the critical temperature of the superconducting ($T/T_C$), for different values of $E_F$, (a) $E'_{F}/\Delta_{0}=100$ and (b) $E'_{F}/\Delta_{0}=1$.}
\end{figure}

Since the thermal conductivity measurements provide information on order parameter symmetry, we now proceed to investigate how the thermal conductance spectra of a N/S borophane change with temperature.
Figure \ref{kth1} shows the normalized thermal conductance, within the relevant temperature region ($0<T<T_C$), as a function of $T/T_C$ for various $E_F$, where (a) $E'_{F}/\Delta_{0}=100$ and (b) $E'_{F}/\Delta_{0}=1$. The influence of Fermi energy can also be seen in this figure. Insets show a zoomed view of the data of the region inside the dashed rectangles. The thermal conductance demonstrates a peak structure in which the peak occur at $T/T_C=0.07$.
This observation is in perfect agreement with the theoretical and experimental expectations of the electronic thermal conductivity in d-wave superconductors ~\cite{R.W.Hill,Zhang-Ong2001}. As pointed out by Hirschfeld {\it et al}. ~\cite{Hirschfeld96}, in the electronic contribution of the thermal conductivity, at least one peak should be appeared below $T_C$.
Here, we propose a mathematical model using Gaussian function, namely the inverse Gaussian function ~\cite{Chhikara}, from the curve fitting which can be expressed as
\begin{eqnarray}\label{bc1}
\kappa(T)=\sqrt{\frac{\lambda}{2\pi T^3}}e^{\frac{-\lambda(T-\mu)^2}{2 T \mu^2}}
\end{eqnarray}
where $\mu>0$ is the mean and $\lambda>0$ is a scaling parameter, making clear that the thermal conductance changes exponentially.
Note that as $\lambda$ tends to infinity, the inverse Gaussian function becomes more like a normal Gaussian function. For the case of $E_{F}/\Delta_{0}=0.01$, the fitting parameters are as $\mu=1$ and $\lambda=1$.
An exponential feature was reported earlier in graphene ~\cite{Yokoyama08} and silicene NIS junctions ~\cite{C.Paul2016}, which is a universal characteristic of the thermal conductance, similar to that of conventional normal metal–superconductor junctions ~\cite{Bezuglyi-PRL-2003}. This exponential fall of thermal conductance with a large peak, when the temperature is below the transition temperature $T_C$, reflects the d-wave symmetry of the borophane superconductor ~\cite{Yokoyama08}.
Similar qualitative feature was found earlier in the case of electronic thermal conductivity in d-wave superconductors~\cite{Lofwande}.

\begin{figure}[t]
\begin{center}
\includegraphics[width=3.3in]{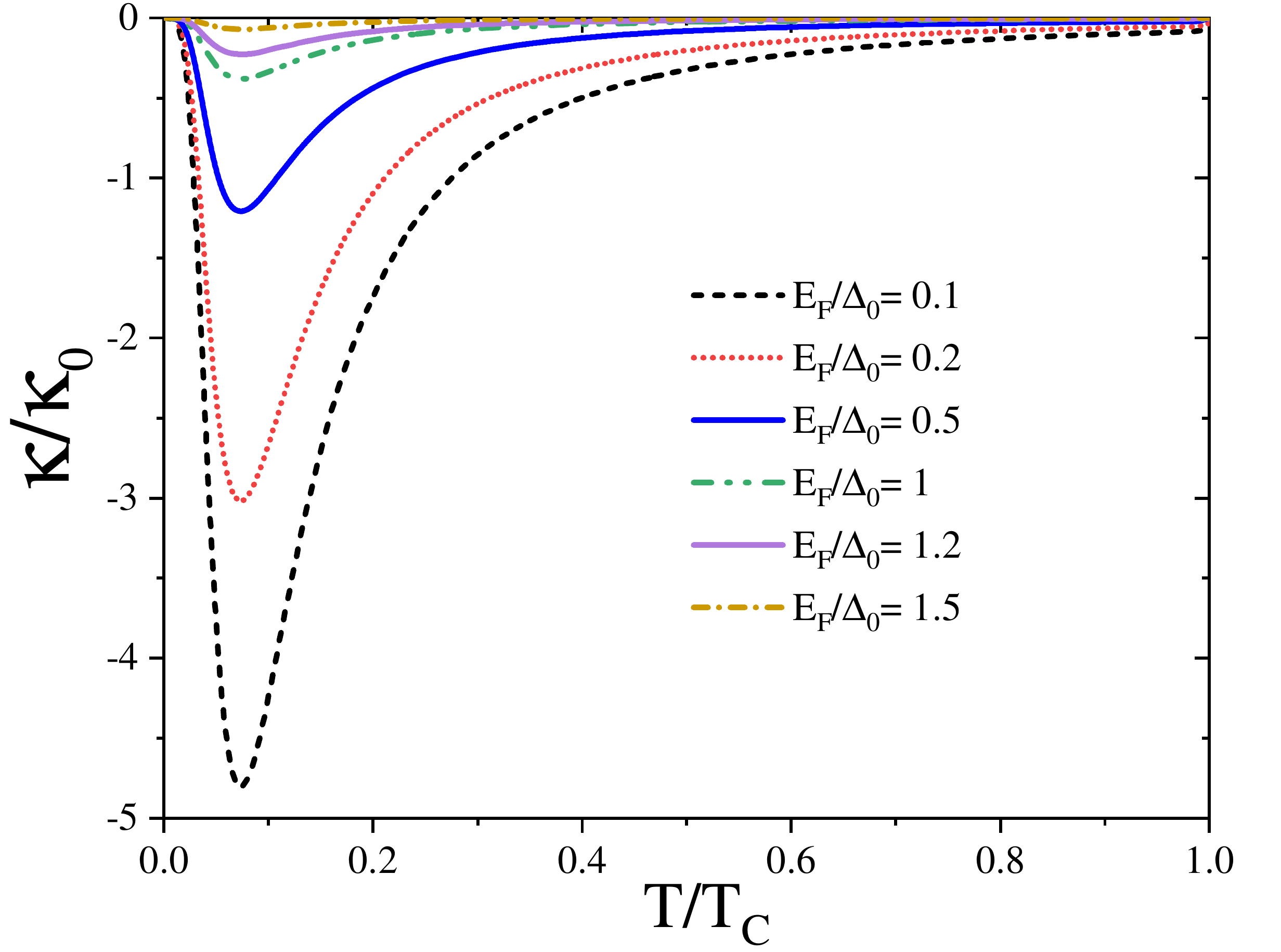}
\end{center}
\caption{\label{kth2} Normalized thermal conductance of the borophane NS junction as a function of the temperature, in units of the critical temperature of the superconducting ($T/T_C$), for different values of $E_F$. Here we set $E'_F/\Delta_0=100$.}
\end{figure}

Figure \ref{kth2} shows the normalized thermal conductance, as a function of $T/T_C$, for the Fermi energy interval of $0.1<E_F<1.5 $ eV for heavily doped superconducting regime, where $E'_F/\Delta_0=100$. As can be seen, the thermal conductance is reduced with increasing the Fermi energy of the normal region. This figure demonstrates pronounced negative differential thermal conductance (NDTC) with a maximum absolute value at $T/T_C=0.07$.
The accessibility of negative differential thermal conductance, i.e., the heat current increases (decreases) as the thermal bias decreases (increases), is an effect that has been widely exploited, as a key building block of thermal circuits. Thus, we can conclude that the most striking feature of the thermal transport in borophane-based NS junction is the observation of the anomalous negative differential thermal conductance~\cite{Li,L.Wang08}.

A number of experiments and theoretical works have already reported the negative differential thermal conductance in graphene nanoribbons~\cite{Vallabhaneni-APL}, topological insulator superconductor junctions~\cite{Ren13}, atomically-thin $WS_2$ field-effect transistors~\cite{Nathawat17} and graphene-based superconducting junction\cite{Zhoua16}. Another interesting implication of NDTC is the effect of thermal bistability ~\cite{Linxiao12}.

\begin{figure}[t]
\begin{center}
\includegraphics[width=3.5in]{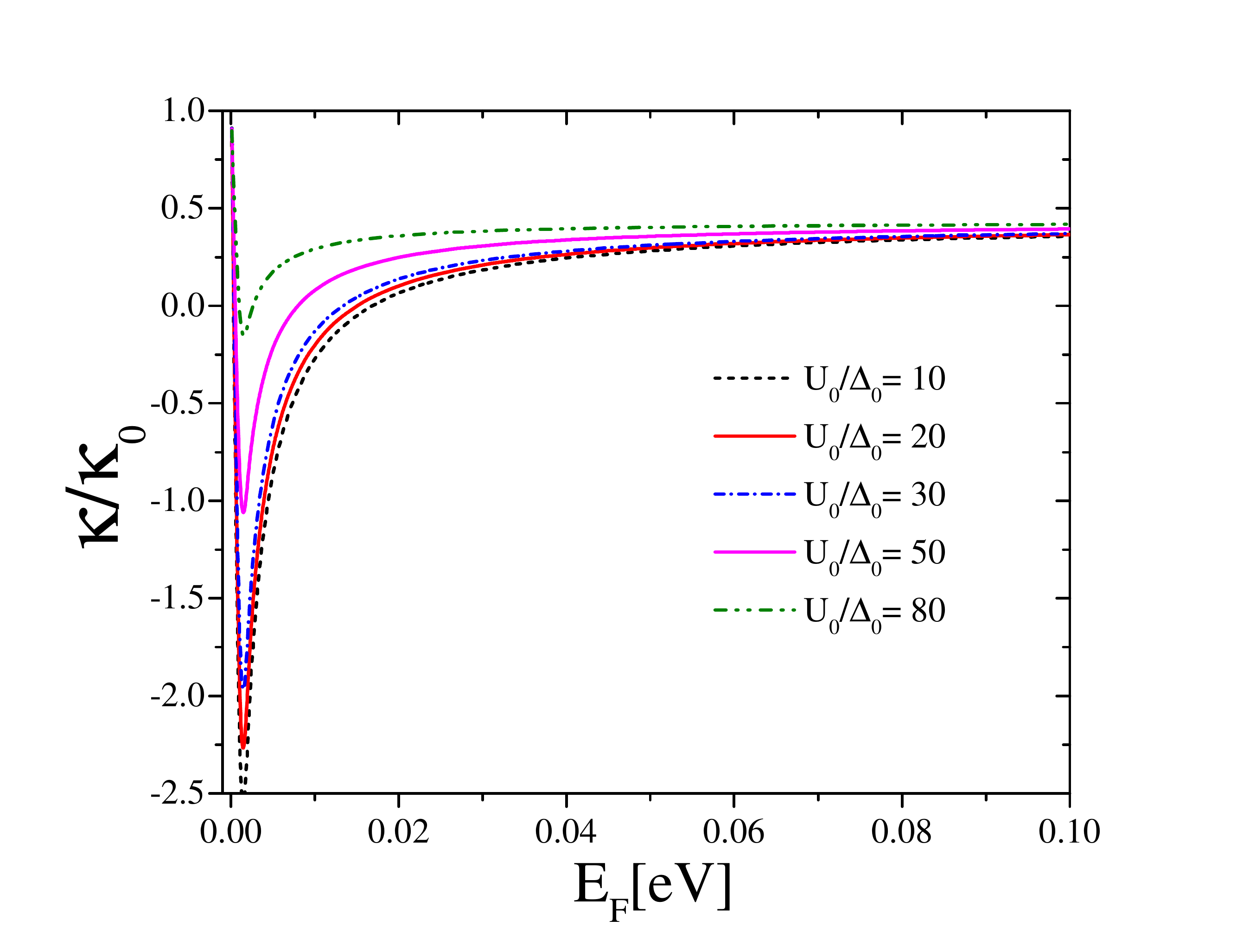}
\end{center}
\caption{\label{kth3} Normalized thermal conductance of the borophane NS junction as a function of the Fermi energy $E_F$, for $T/T_C=0.1$ for different values of $U_0$, where $E'_F/\Delta_0=100$.}
\end{figure}

Finally, to gain more insight into the thermal rectification effects in the borophane-based NS junction profoundly, displayed in Fig.\ref{kth3}, we plot the differential thermal conductance versus the Fermi energy at $T/T_C=0.1$, for different values of $U_0$ with $E'_F/\Delta_0=100$.
Interestingly, this figure shows that the sign of the DTC, with a maximum absolute value at $E_F\sim$2 meV, is tunable by Fermi energy control of the normal borophane region. It is evident that the larger the $U_0$, the smaller peak of the thermal conductance.
Remarkably, this ability to tune and control NDTC by Fermi energy, provides potential ways to manage heat and manipulate thermal signals thermal transistors ~\cite{Li} and thermal logic~\cite{L.Wang08}.

\section{Conclusion}\label{sec:concl}

In conclusion, we have studied the charge and heat transport in a normal-metal/superconductor (NS) junction of the tilted anisotropic Dirac cone material borophane, using the extended Blonder-Tinkham-Klapwijk formalism. The conductance spectra of NS borophane, a two-dimensional Dirac semimetal with two tilted anisotropic Dirac cones in its dispersion, is investigated. Completely different from the usual normal-metal-superconductor junctions, in spite of the large mismatch in Fermi wavevectores of the normal-metal and superconductor sides of the borophane NS junction, the electron-hole conversion happens with unit probability at normal incidences.
Furthermore, we demonstrate that in heavily doped superconducting regime, for sufficiently large absolute values of the Fermi energy at the normal side ($E_F/\Delta_{0}\gg1$), the electron-hole conversion happens with unit probability, almost at any incident angle.
Interestingly, a zero-bias conductance peaks appears in the conductance spectra. This two peak structure is just similar to the case of normal metal—insulator d-wave superconductor junction.
Exactly the same as unconventional anisotropic d-wave superconductor-graphene junctions, the subgap conductance is always close to $2G_0$, but becomes more constant with increasing $E_{F}$.
The dependence of the Andreev conductance on the Fermi energy and bias voltage, enable us selecting the retro configuration or specular configuration in types of Andreev reflection processes.
From the curve fitting, we numerically find that independent of the Fermi energy, the temperature dependence of the differential thermal conductance in borophane can be modelled as an inverse Gaussian function, reflecting the d-wave symmetry of the borophane superconductor.
We propose a scheme for achieving negative differential thermal conductance in borophane NS junction, as a key building block of thermal circuits.
The most striking feature of the anomalous thermal transport in borophane-based NS junction, described in this paper, is the observation of the negative differential thermal conductance.The sign of the DTC, is tunable
It has been found that the switch between positive and negative differential thermal conductance can be controlled
electrically, by Fermi energy control of the normal borophane region.
Our findings will have potential applications in developing borophane-based thermal management and signal manipulation mesoscopic structures such as heat transistors, heat diodes and thermal logic gates.

\section{acknowledgments}
This work was partially supported by Iran Science Elites Federation.

\end{document}